# Rough Interfaces Beyond the Gaussian Approximation

M. Caselle, F. Gliozzi, S. Vinti[a], R. Fiore[b], M. Hasenbusch[c], and K. Pinn[d] *

[a]Dipartimento di Fisica Teorica, Università di Torino, Via P. Giuria 1, I-10125 Torino, Italy

[b]Dipartimento di Fisica Teorica, Università della Calabria, I-87030 Cosenza, Italy

[c]DAMTP, Silver Street, Cambridge, CB3 9EW, England

[d]Institut für Theoretische Physik I, Universität Münster, Wilhelm-Klemm-Str. 9, D-48149 Münster

We compare predictions of the Capillary Wave Model with Monte Carlo results for the energy gap and the interface energy of the 3D Ising model in the scaling region. Our study reveals that the finite size effects of these quantities are well described by the Capillary Wave Model, expanded to two-loop order (one order beyond the Gaussian approximation).

An effective model widely used to describe a rough (fluid) interface is the Capillary Wave Model (CWM). In its simplest formulation one assumes an effective Hamiltonian proportional to the variation of the interface's area with respect to the classical solution. Until recently [1] the CWM has been studied only in its quadratic approximation. The improvement of Monte Carlo (MC) calculations reached in the last years now allows tests of the CWM beyond the Gaussian level. This paper is a very short version of ref. [2]. We compare 2-loop predictions of the CWM for the tunneling mass gap and the interface energy with MC results for a rough interface in the scaling region of the 3D Ising model.

## 1. THE MODELS

### 1.1. The Ising Model

We consider the 3D Ising model on cubic lattices of size $L_1, L_2$ in the $x-$, $y-$directions, with $L_2 \geq L_1$, and of size $t$ in $z$-direction. In the $x-$, $y-$direction we always have periodic boundary conditions, whereas in $z$-direction either periodic or antiperiodic boundary conditions are used. The Hamiltonian is a sum over nearest neighbor pairs, $H = -\sum_{<i,j>} s_i s_j$, and the Boltzmannian is $\exp(-\beta H)$. While in infinite volume, for $\beta > \beta_c$, the system shows a spontaneous symmetry breaking, in finite volume this cannot occur, and interfaces spontaneously appear, separating domains of different magnetization. The critical coupling of the model is $\beta_c = 0.221652(3)$ [3]. In the region between the critical and the roughening coupling $\beta_r = 0.4074(3)$ [4] the interface behaves essentially as a 2D critical system. It is a widely anticipated assumption that this system can be described by a (continuum) model with a Hamiltonian that is proportional to the area of the interface.

### 1.2. The Capillary Wave Model

If the interface is described by a single valued height function $\phi$ over the torus with coordinates $0 \leq x_i \leq L_i$, $i = 1, 2$, the CWM Hamiltonian is $\mathcal{H}(\phi) = \mathcal{A}(\phi) - L_1 L_2$, with

$$\mathcal{A}[\phi] = \int_0^{L_1} dx \int_0^{L_2} dy \sqrt{1 + \left(\frac{\partial \phi}{\partial x}\right)^2 + \left(\frac{\partial \phi}{\partial y}\right)^2}. \quad (1)$$

We are interested in the free energy $F$, defined through

$$\begin{aligned} F &= -k_B T \ln Z \\ Z &= \lambda \, \exp(-\sigma L_1 L_2) \, Z_q(\sigma, L_1, L_2) \\ Z_q &= \int [D\phi] \exp(-\mathcal{H}[\phi]) . \end{aligned} \quad (2)$$

$\lambda$ is an undetermined constant in this approach. By expanding the root in eq. (1), one can set up

---
*Speaker at the conference

a perturbative expansion for $Z_q$. The expansion parameter is $(\sigma A)^{-1}$, with $A = L_1 L_2$. Unfortunately, the model is UV (by naive power counting) nonrenormalizable at large momenta. However, at least to 2-loop, the same finite results are obtained by a large class of regularizations [5,2,6]. To 2-loop order, one gets $Z_q = Z_q^{(1l)} Z_q^{(2l)}$, with

$$Z_q^{(1l)}(u) = \frac{1}{\sqrt{u}} \left| \eta(iu)/\eta(i) \right|^{-2}. \tag{3}$$

$\eta$ is the Dedekind eta function, and $u = L_2/L_1$ is the asymmetry parameter. For the 2-loop contribution one obtains

$$Z_q^{(2l)} = 1 + f(u)(\sigma A)^{-1}, \tag{4}$$

with

$$f(u) = \frac{1}{2} \left\{ \left[ \frac{\pi}{6} u E_2(iu) \right]^2 - \frac{\pi}{6} u E_2(iu) + \frac{3}{4} \right\}. \tag{5}$$

Here, $E_2$ denotes the first Eisenstein series. Corrections from higher loops are of order $(\sigma A)^{-2}$.

## 2. OBSERVABLES AND MC

The observables that we used to check predictions of the CWM were the *energy gap* $\Delta E$ and the *interface energy* $E_S$.

Due to tunneling, there is a small energy gap $\Delta E$ between the symmetric and antisymmetric ground state of the Ising model in a finite geometry. In the dilute gas approximation, the energy splitting is directly proportional to the interface partition function, $\Delta E = Z(\sigma, L_1, L_2)$, and thus allows for a comparison with the CWM partition function. We computed the energy gap for several values of $\beta$ and for various combinations of $L_1$ and $L_2$ employing two different methods: the time slice correlations method (TSC) (see, e.g. [7]) and the boundary flip method (BF) [8].

Our second observable was the interface energy, that can be computed from the CWM partition function, $E_S = -Z^{-1}(\partial Z/\partial \beta)$. The 2-loop result is

$$E_S(\sigma, L_1, L_2) = \sigma' L_1 L_2 - \frac{\lambda'}{\lambda} - \frac{1}{Z_q^{(2l)}} \partial_\beta Z_q^{(2l)}. \tag{6}$$

Here, $\sigma' \equiv \partial_\beta \sigma$ and $\lambda' \equiv \partial_\beta \lambda$.

For the Ising model, we estimated the interface energy from the difference $E_S = \langle H \rangle_p - \langle H \rangle_a$ of the energy expectation value with periodic and antiperiodic boundary conditions, respectively. These expectation values were computed using a microcanonical demon algorithm in combination with a particularly efficient canonical update of the demons [9]. The algorithm was implemented using the multi-spin coding technique.

## 3. MC DATA ANALYSIS

### 3.1. Energy Gap

Our $\beta$-values are listed in table 3.1. For these values, the infinite volume bulk correlation length ranges from $\xi \approx 2.6$ for $\beta = 0.2275$ to $\xi \approx 4.5$ for $\beta = 0.2240$. We simulated for typically 10 to 20 different combinations of $L_1$ and $L_2$, where $L_2 \geq L_1$. We always kept $L_1$ larger than the inverse deconfinement temperature of the dual gauge model. $t$ was usually fixed to 120, for simulations with very large $L_1$ and $L_2$ also larger. To test the CWM prediction, we fitted the 2-loop result to our MC data. Note that there are only the two parameters $\lambda$ and $\sigma$ to be fitted.

Table 1
$\beta$-*values for* $\Delta E$-*measurements. T means TSC method, B means BF method*

| $\beta$ | $L_1^{min}$ |   | $\beta$ | $L_1^{min}$ |     |
|---------|-------------|---|---------|-------------|-----|
| 0.2240  | 18          | B | 0.2258  | 12          | T   |
| 0.2246  | 13          | T | 0.2275  | 10          | T,B |

As an example, we show in fig. 1 our MC results for three of the $\beta$-values, together with the best fit curves. Obviously, the size and shape dependence of $\Delta E$ is nicely described by the 2-parameter fit with the 2-loop CWM partition function.

Compared to fits with the classical result, the reduced $\chi^2$ of the fits with the 2-loop-result were always more than one order of magnitude smaller.

### 3.2. Interface Energy

For the interface energy, the 2-loop contribution appears as an additive correction to the

Gaussian (and classical) result. We made our calculations at $\beta = 0.24$. Precise estimates of $\sigma$ and $\sigma'$ were known from other MC experiments. We considered these estimates as external input. Let us define the following energy differences:

$$\Delta E_S(L) = E_S(2L, L/2) - E_S(L, L) \qquad (7)$$

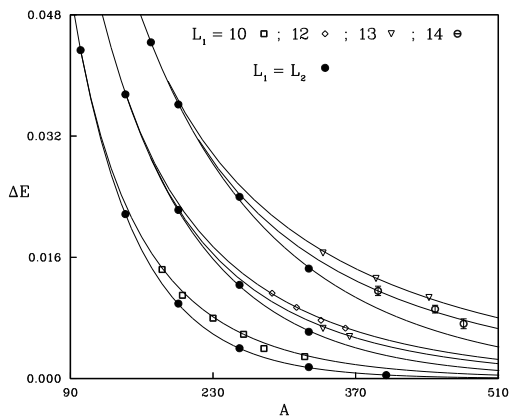

Fig 1: MC data and best 2-loop fits for different values of $\beta$ versus $A = L_1 L_2$. From top to bottom, the first three lines correspond to $\beta = 0.2246$, the next three to $\beta = 0.2258$, and the last two to $\beta = 0.2275$.

The CWM prediction for this quantity is

$$\Delta E_S^{(CWM)} = \frac{\sigma'}{\sigma^2 L^2} [f(4) - f(1)]. \qquad (8)$$

In fig. 2, we plot this prediction together with the corresponding estimates obtained from our MC data for the interface energy. A very nice agreement is seen for $L \geq 16$.

## CONCLUSION

Our results show that the finite size effects of interface properties in the scaling region of the Ising model are well described by the 2-loop expanded CWM.

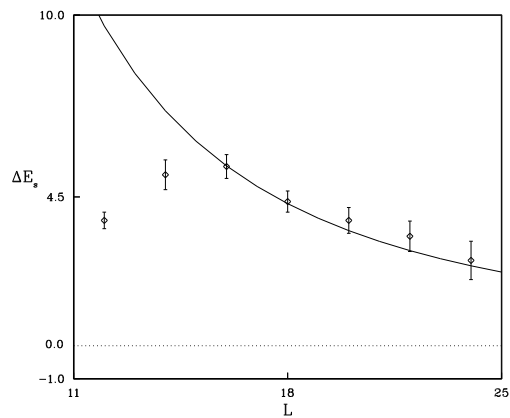

Fig 2: Comparison between MC and CWM predictions for $\Delta E_S$. The dotted line is the theoretical prediction if 2-loop corrections are neglected.